\documentclass[aps,prc,preprint,groupedaddress,showpacs]{revtex4}
\usepackage{youngtab,color}
\usepackage{multirow}
\usepackage{epsfig}
\usepackage{longtable}
\begin{document}

\title{Spectrum of low-lying $s^{3}Q\bar{Q}$ configurations with negative parity}

\author{S. G. Yuan$^{1,2,3,4}$}
\author{C. S. An$^{2,3}$}\email{ancs@ihep.ac.cn}
\author{K. W. Wei$^{2,3,5}$}
\author{B. S. Zou$^{2,3,6}$}\email{zoubs@ihep.ac.cn}
\author{H. S. Xu$^{1,3}$}
\affiliation{1. Institute of Modern Physics, Chinese Academy of
Sciences, Lanzhou 730000, China \\ 2.Institute of High Energy
Physics, Chinese Academy of Sciences, Beijing 100049, China\\ 3.
Theoretical Physics Center for Science Facilities, Chinese Academy
of Sciences, Beijing 100049,
China\\
4. Shanghai Institute of Applied Physics, Chinese Academy of Sciences, 
Shanghai 201800, China\\
5. College of Physics and Electrical Engineering, Anyang Normal
University, Henan 455002, China
\\ 6.  State Key Laboratory of
Theoretical Physics, Institute of Theoretical Physics, Chinese
Academy of Sciences, Beijing 100190, China }

\thispagestyle{empty}

\date{\today}

\begin{abstract}

Spectrum of low-lying five-quark configurations with strangeness
quantum number $S=-3$ and negative parity is studied in three kinds
of constituent quark models, namely the one gluon exchange,
Goldstone Boson exchange, and instanton-induced hyperfine
interaction models, respectively. Our numerical results show that
the lowest energy states in all the three employed models are lying
at $\sim$1800 MeV, about 200 MeV lower than predictions of various
quenched three-quark models. In addition, it is very interesting
that the state with the lowest energy in one gluon exchange model is
with spin $3/2$, but $1/2$ in the other two models.

\end{abstract}

%%% 12.39.-x    Phenomenological quark models
%%% 14.20.Jn    Hyperons
%%% 14.20.Pt    Exotic baryons

\maketitle

\section{introduction}
\label{sec:intro}

Up to now, the experimental information on the $\Omega$ hyperon
excitations is still very poor~\cite{pdg}. There are only four
$\Omega$ states
discovered~\cite{barns,Biagi:1985rn,Aston:1987bb,Aston:1988yn},
namely the ground state $\Omega(1672)$, and three excitations
$\Omega(2250)$, $\Omega(2380)$ and $\Omega(2470)$. Among these four
states, only $\Omega(1672)$ has been clarified to have spin parity
quantum number $\frac{3}{2}^{+}$, while quantum numbers for the
other three have not been pinned down yet.

Although there has not been any further experimental evidence about
$\Omega^{*}$ since 1990s, theorists are always interested in
spectrum of $\Omega$ excitations, which has been investigated within
various approaches, such as constituent quark models based on one
gloun exchange~\cite{CI86,CIK81}, Goldstone Boson
exchange~\cite{GR96b,Helminen} and instanton-induced~\cite{metsch2}
interactions, respectively, large $1/N_{c}$ expansion of
QCD~\cite{CC00,SGS02,GSS03,MS04b,MS06b}, bound state approach of
Skyrme model~\cite{oh}, meson-baryon chiral dynamics~\cite{oset},
etc. Most of the theorists have tried to compare their predictions
with data, while information from experiments is so poor that
definite comparison seems to be impossible. However, to give more
guidance for future experimental measurements, systematical studies
of $\Omega^{*}$ spectrum in different theoretical approaches are
valuable.

Recently, the traditional constituent quark model is developed to
include higher Fock components, in this extended quark model,
wave function of a baryon can be expressed as
\begin{equation}
 |\psi\rangle_{B}=\frac{1}{\mathcal{\sqrt{N}}}\left(|QQQ\rangle+
 \sum_{i,n_{r},l}C_{in_{r}l}|QQQ(Q\bar{Q}),i,n_{r},l\rangle \right)\,,
\label{wfn}
\end{equation}
where the first term is the conventional wave function for the baryon with three
constituent quarks, and the second term is a sum over all possible higher Fock
components with a $Q\bar{Q}$ pair.
Different possible five-quark components are distinguished by inner radial and 
orbital quantum numbers $n_{r}$ and $l$, and $i$ numbers different five-quark
configurations with same $n_{r}$ and $l$.
Finally, $C_{in_{r}l}/\sqrt{\mathcal{N}}\equiv A_{in_{r}l}$
represents the probability amplitude for the corresponding five-quark component, and
$C_{in_{r}l}$ can be calculated by 
$C_{in_{r}i}=\langle QQQ\arrowvert\hat{T}\arrowvert QQQ(Q\bar{Q}),i,n_{r},l\rangle/(M_{B}-E_{in_{r}l})$,
once we make choice of the transition mechanism between 
three- and five-quark components in baryon $B$ with mass $M_{B}$.
It is shown that there should be
notable five-quark components in the baryon resonances
\cite{Bijker:2009up,Santopinto:2010zza,Li:2005jb,Li:2005jn,Li:2006nm,JuliaDiaz:2006av,An:2008tz,An:2008xk,An:2010wb,An:2011sb}.
So it immediately occurs to us that the five-quark components in
$\Omega^{*}$ should be very important, because all the three valence
quarks in $\Omega^{*}$ are strange ones, then the five-quark
configurations with $q\bar{q}$ pairs (we denote $Q\bar{Q}\equiv
q\bar{q}$ and $Q\bar{Q}\equiv s\bar{s}$ for light and strange
quark-antiquark pairs, respectively.) must play very special roles
in the properties of $\Omega^{*}$. Since the $q\bar{q}$ pair has
different flavor with valence strange quarks, it has two advantages
for the study of five-quark components. First, there is not any
Pauli blocking effect for the $q$, which would result in lower
excitation energy for the new excitation mechanism of $\Omega$
states by pulling out a $q\bar q$ pair and make the five-quark
components larger in $\Omega^*$ than in $N^*$, $\Lambda^*$, etc.
Secondly it simplifies the model calculation of the five quark
system.

%In fact, one may anticipate that five-quark configurations with
%$q\bar{q}$ are the dominant components in physical $\Omega^{*}$
%states, since pulling out a $Q\bar{Q}$ pair from the quark sea that
%results in a five-quark state can be treated as a way to excite the
%$\Omega$ hyperon with three quarks, and corresponding energies for
%this kind of excitations should be lower than those for traditional
%orbital excitations of quarks, accordingly, five-quark components
%may be more preferable $\Omega^{*}$.

To specify the roles of $s^{3}Q\bar{Q}$ in $\Omega^{*}$,
as the first step, we investigate the spectrum of low lying
$s^{3}Q\bar{Q}$ configurations with negative parity in this manuscript,
employing the constituent quark model complemented by three
different kinds of hyperfine interactions between quarks, i.e. the one
gluon exchange (OGE), Goldstone Boson exchange (GBE) and instanton-induced (INS) force
interactions, respectively. In order to get more accurate
information about the five-quark configurations with
light $q\bar{q}$ pairs, the $SU(3)\otimes SU(2)\otimes O(3)$
breaking corrections are taken into account in our model.

The present manuscript is organized as follows. In Section \ref{sec:frame},
we present our theoretical framework, which includes explicit
forms of the employed hyperfine interactions between quarks.
Numerical results for spectrum
of the studied five-quark configurations in our model are
shown in Section \ref{sec:result}. Finally, Section \ref{sec:end}
contains a brief conclusion.

\section{Theoretical Framework}
\label{sec:frame}

In the constituent quark model, Hamiltonian for a five-quark
system with a $Q\bar{Q}$ pair can be expressed as:
\begin{eqnarray}
H&=&H_{o}+H_{hyp}+\sum_{i=1}^{5}m_{i}\,,
\label{ham}
\end{eqnarray}
where $m_{i}$ denote the constituent masses of the
quarks, $H_{hyp}$ is the
hyperfine interaction between quarks, which is often
treated as perturbation, and $H_{o}$ the Hamiltonian
concerning orbital motions of the quarks which
should contain two parts, namely the kinetic term and
confining potential of the quarks. Both of the orbital
Hamiltonian $H_{o}$ and the hyperfine interaction $H_{hyp}$
for three-quark system have been intensively discussed
in literatures, here we will develop them to
five-quark system explicitly
in Sec.~\ref{sec:orb} and~\ref{sec:hyp}.

\subsection{The orbital Hamiltonian for a five-quark system}
\label{sec:orb}

In this manuscript, the employed form of $H_{o}$ is as follow
\begin{equation}
H_{o}=\sum_{i=1}^5 {\vec{p}_i^2\over 2 m_{i}}+\sum_{i<j}^5
V_{conf}(r_{ij})\,,
\label{ho}
\end{equation}
where $\vec{p}_{i}$ and $m_{i}$ denote the momentum and mass
of the ith quark, and $V_{conf}(r_{ij})$ is the quark confinement 
potential. The harmonic oscillator, as one of the most commonly used
quark confinement potentials, has been successfully applied 
to the spectroscopy of nonstrange and strange baryon excitations~\cite{CI86,GR96b}.
On the other hand, in the present case, the studied states are the low-lying 
$s^{3}Q\bar{Q}$ configurations with negative parity, namely, 
both the inner quantum numbers $i$ and $n_{r}$ in Eq.~(\ref{wfn}) are $0$, 
consequently, here we just take $V_{conf}(r_{ij})$ to be the harmonic 
oscillator form as follow
\begin{equation}
V_{conf}(r_{ij})=-\frac{3}{8}\lambda_i^C\cdot\lambda_j^C\left[C(\vec{r}_i-\vec{r}_j)^2+V_0\right].\label{conf}
\end{equation}
Here $\lambda_{i(j)}^{C}$ are the Gell-Mann matrices in the $SU(3)$ color space,
$C$ denotes a confinement strength constant, and $V_{0}$ represents
the unharmonic part of $V_{conf}$ which is treated as a constant in this manuscript.
Explicit calculations show that the
matrix elements of $\lambda_{i}^{C}\cdot\lambda_{j}^{C}$
between both of $QQ$ and $Q\bar{Q}$ pairs in a five-quark system are the same value $-4/3$,
which is half of the matrix element between a $QQ$ pair in a three-quark system.
On the other hand, the $Q\bar{Q}$ pair in $s^{3}Q\bar{Q}$ system 
can be either light $q\bar{q}$ or $s\bar{s}$. Considering the
difference of constituent masses of the light and strange quarks,
one has to carefully treat $H_{o}$ for $s^{4}\bar{s}$ and $s^{3}q\bar{q}$.
In general, we can take the constituent masses of 
light quarks to be same as that of the strange quark,
and treat corrections from different masses
as perturbation. Consequently, $H_{o}$ can be rewritten as the following form
\begin{equation}
 H_{o}=\sum_{i=1}^{4}\left(\vec{\tilde{p}}_{i}^{2}/2{m}_{s}
 +\frac{5C}{2}\vec{\xi}_{i}^{2}\right)+5V_{0}+H_{o}^{\prime}\,,
\label{ho'}
\end{equation}
where $\xi_{i}$ represent the Jacobi coordinates for a system of $5$ constituents
which are defined as follow
\begin{eqnarray}
 \vec{\xi_{i}}&=&\frac{1}{\sqrt{i+i^{2}}}\left(\sum_{j=1}^{i}\vec{r}_{j}-i\vec{r}_{i}\right),~i=1,\cdots,4\,,\nonumber\\
 \vec{R}_{cm}&=&\frac{1}{5}\sum_{i=1}^{5}\vec{r}_{i}\,,
 \label{jac}
\end{eqnarray}
and $\vec{\tilde{p}}_{i}$ the corresponding momentums.
$H_{o}^{\prime}$ denotes corrections from different quark masses,
in the case of $Q\bar{Q}=s\bar{s}$, $H_{o}^{\prime}$ is zero, and 
we will discuss the $Q\bar{Q}=q\bar{q}$ case at the end of this section.

If we neglect contributions from the perturbation term $H_{o}^{\prime}$,
the eigenvalue of
$H_{o}$ should be
\begin{equation}
 E_{o}=(N+6)\omega+5V_{0}\,,
 \label{eng}
\end{equation}
where $\omega$ denotes the oscillator frequency
which is defined as $\omega=\sqrt{5C/m_{s}}$
, and $N=2N^{r}+L$ is the number
of excited quanta of the oscillators.
In this manuscript, we consider
the $s^{3}Q\bar{Q}$ configurations with $N^{r}=L=0$, namely all
the quarks and antiquarks in the five-quark configurations are in their ground
states, then the energy~(\ref{eng}) reduces to $E_{o}=6\omega+5V_{0}$.
$\omega$ can be determined in the following way, as we know,
the oscillator frequency $\omega_{3}$ for a
three light quarks system is $\omega_{3}=\sqrt{6C/m}$
with $m$ the constituent mass of the light quark,
then we can obtain that $\omega=\sqrt{\frac{5m}{6m_{s}}}\omega_{3}$.
And $\omega_{3}$ can be inferred from the
empirical radius of the proton which leads
to $\sqrt{m\omega_{3}}\simeq246$ MeV, or mass splitting
between nucleon and $N(1440)$ which yields $\omega_{3}=251$ MeV.
We take the latter value in this manuscript.

And the eigenfunction in the case of $N=0$
is just a completely symmetric wave function regarding the four
Jaccobi coordinates, then we can construct the complete wave functions
of $s^{3}Q\bar{Q}$ configurations in the following way:
first, color wave functions of the four-quark subsystems
in any $QQQQ\bar{Q}$ systems must be $[211]_{C}$, to
combine with the anti-quark to form a color singlet,
consequently, the flavor-spin symmetry
of the four-quark subsystems in the $s^{3}Q\bar{Q}$ configurations
can only be $[31]_{FS}$. Moreover, the flavor wave functions
of a $sssq$ system can be completely symmetric $[4]_{F}$ or
mixed symmetric $[31]_{F}$. As shown in~\cite{Helminen},
four kinds of four-quark wave functions can meet
the above requirements. Coupling these four-quark
systems to the antiquark, one can obtain two five-quark
configurations with $s\bar{s}$, and seven configurations
with $q\bar{q}$. The total spin for these configurations
can be $1/2$, $3/2$ and $5/2$, as shown in Table~\ref{con}.
Explicit wave functions for these configurations
can be easily derived from the ones given in~\cite{An:2005cj}.

Finally, in the case of $Q\bar{Q}$ being a light quark-antiquark pair,
we have to take into account the perturbation term $H_{o}^{\prime}$.
As shown in~\cite{Glozman:1995xy}, we can take this term to be a
flavor-dependent one as follow
\begin{equation}
H_{o}^{\prime}=\frac{m_{s}-m}{m_{s}}\sum_{i=1}^{4}
\frac{\vec{p}_{i}^{2}+\vec{p}_{5}^{2}}{2m}
\delta_{iq}\,,
\end{equation}
here $m$ is the constituent mass of the light quark, $\delta_{iq}$
is a flavor dependent operator with eigenvalue $1$ for light quark and $0$
for strange quark.
In the present case, all of the quarks and antiquark are in their
ground states, one can get that the matrix elements of $H_{o}^{'}$
between all of the studied configurations are the same one
\begin{equation}
 \langle H_{o}^{'} \rangle=\frac{6}{5}(\frac{m_{s}}{m}-1)\omega\,,
 \label{mcorrt}
\end{equation}

\subsection{The hyperfine interaction between quarks}
\label{sec:hyp}

As stated in Sec~\ref{sec:orb}, there are nine different
configurations studied in this manuscript. If the approximation
that $H_{hyp}=0$ is applied, the five-quark configurations with $q\bar{q}$
and $s\bar{s}$ should be two categories of degenerate states.
To calculate the mass splittings of the degenerate configurations,
explicit perturbative hyperfine interactions
are needed. Here we employ
three different kinds of spin-spin forces, which are mediated by
one gluon exchange~\cite{Isgur:1979be,Isgur:1978wd,Isgur:1977ef,Capstick:2000qj},
Goldstone Boson exchange~\cite{GR96b,Helminen} and instanton-induced
interactions~\cite{metsch1,metsch2,Blask:1990ez,Klempt:1995ku,Koll:2000ke}, respectively,
as hyperfine interactions between quarks.
These interactions between both of $QQ$ and $Q\bar{Q}$ pairs have
been explicitly given in literatures, here we just present a very brief review,
and apply them to a five-quark system.

In the OGE model,
hyperfine interaction between quarks can be expressed
as chromomagnetic form, which has been intensively used
in the studies of multiquark
configurations \cite{Leandri:1989su,SilvestreBrac:1992yg,Buccella:2006fn,Hogaasen:2005jv}.
Here we employ the form given in~\cite{Buccella:2006fn,Hogaasen:2005jv}
\begin{equation}
H_{hyp}^{OGE}=-\sum_{i,j}C_{i,j}\vec\lambda_i^{C}\cdot
\vec\lambda_j^{C} \vec\sigma_i\cdot\vec\sigma_j, \label{cs}
\end{equation}
where $\vec\sigma_i$ are the Pauli spin matrices, $\vec\lambda_i^c$ the
Gell-Mann $SU(3)_C$ color matrices, and $C_{i,j}$ a flavor dependent chromomagnetic
interaction strength operator. If the subscript $i$ or $j$
indicates an antiquark, the following
replacement should be applied:
$\vec\lambda^{C}\rightarrow -\vec\lambda^{C*}$.
Once the overall symmetry is taken into account, the hyperfine
interaction~(\ref{cs}) should reduce to
\begin{equation}
H_{hyp}^{OGE}=-6C_{1,2}\vec\lambda_1^{C}\cdot
\vec\lambda_2^{C} \vec\sigma_1\cdot\vec\sigma_2
+4C_{4,5}\vec\lambda_4^{C}\cdot
\vec\lambda_5^{C*} \vec\sigma_4\cdot\vec\sigma_5
\end{equation}
The resulting matrix elements of the color, spin and flavor dependent chromomagnetic strength
operators are given explicitly in Appendix~\ref{app:oge}.
And values of interaction strength constants $C_{QQ}$ and $C_{\bar{Q}Q}$ can be found
in~\cite{Buccella:2006fn,Hogaasen:2005jv,Yuan:2012wz}.

In Ref.~\cite{Helminen}, the GBE model is applied to calculate spectrum of the
five-quark configurations with quantum numbers of $N$,
$\Lambda$, $\Sigma$, $\Xi$ and $\Omega$ resonances. While in that paper,
the hyperfine interaction is of
a $SU(3)$ symmetric form, namely, the coupling strength
of $\pi$, $K$ and $\eta$ meson exchanges are taken to
be a same one. In present case, we take the $SU(3)$ broken
form of $H_{hyp}^{GBE}$ which is given in~\cite{GR96b}
\begin{equation}
H_{hyp}^{GBE}=-\sum_{i,j}^{4}C_{i,j}^{M}\vec\lambda_i^{F}\cdot
\vec\lambda_j^{F} \vec\sigma_i\cdot\vec\sigma_j, \label{fs}
\end{equation}
where $\lambda_i^F$ are the
Gell-Mann matrices in flavor space, and $C_{i,j}^{M}$ a
flavor dependent operator for strength of a meson $M$ exchange between
the $ith$ and $jth$ quarks. Notice that in the GBE model~\cite{GR96b},
hyperfine interaction between quark and antiquark
is assumed to be automatically included in the GBE interaction,
so the spin-spin interaction $H_{hyp}^{GBE}$ in Eq.~(\ref{fs}) is
restricted to the four-quark subsystem. After taking into account
the overall symmetry, $H_{hyp}^{GBE}$ reduces to linear combination
of interaction between the first two quarks and that between the fourth
quark and the antiquark:
\begin{equation}
 H_{hyp}^{GBE}=-6C_{1,2}^{M}\vec{\lambda}_{1}^{F}\cdot\vec{\lambda}_{2}^{F}
\vec{\sigma}_{1}\cdot\vec{\sigma}_{2}\,,
\end{equation}
the matrix elements of the spin operator $\vec{\sigma}_{1}\cdot\vec{\sigma}_{2}$
are the same as those in the OGE model which are given in Appendix~\ref{app:oge},
and we list the matrix elements of the flavor dependent operator
$C_{1,2}^{M}\vec{\lambda}_{1}^{F}\cdot\vec{\lambda}_{2}^{F}$ in Appendix~\ref{app:gbe}.
Notice that we have to treat properly the three subsets of the $\eta$ meson
exchange between two quarks, namely, the exchange interaction
for pairs of light quarks($C^{u\bar{u}}$~and~$C^{d\bar{d}}$),
and two strange quarks ($C^{s\bar{s}}$). It's proposed by Glozman and Riska
in~\cite{GR96b} that the strengths of the meson exchange interactions
may have the following relations
\begin{equation}
 C^{u\bar{u}}=C^{d\bar{d}}=C^{\pi};~C^{K}=\frac{m}{m_{s}}C^{\pi};~
 C^{s\bar{s}}=(\frac{m}{m_{s}})^{2}C^{\pi}\,,
\end{equation}
and the phenomenological value for $C^{\pi}$ is around
20$\sim $30 MeV~\cite{Helminen,GR96b}.

Finally, after taking into account the overall symmetry as
what we have done for the OGE and GBE interactions,
the instanton-induced hyperfine interaction between quarks is of the
form~\cite{Blask:1990ez,Klempt:1995ku,Koll:2000ke}
\begin{eqnarray}
H_{hyp}^{INS}&=&-12\left(gP_{1,2}^{qq}+g^{\prime}P_{1,2}^{qs}\right)
\left(P^{S=1}_{1,2}P^{C,6}_{1,2}+2P^{S=0}_{1,2}P^{C,\bar 3}_{1,2}\right)\nonumber\\
&&+4\hat g_{4,5}\left[\frac32P^{S=1}_{4,5}P^{C,8}_{4,5}+P^{S=0}_{4,5}
\left(\frac12P^{C,8}_{4,5}+8P^{C,1}_{4,5}\right)\right]\,,
\end{eqnarray}
where the first term is the operator for interaction
between two quarks, and the second for quark-antiquark
interaction, $P^{QQ'}$ are the projectors on flavor antisymmetrical
$QQ^{\prime}$ states, $P^{S=x}$ the projectors on spin $x$ states,
and $P^{C,y}$ the projectors on color representation of dimension $y$,
respectively. $g$ and $g^{\prime}$ are the strength constants
of interactions between two light quarks, and one light quark
and one strange quark, respectively.
Finally, $\hat g_{4,5}$ is a flavor projector operator, its eigenvalue is
-$g$ for $q \bar q$ and -$g^{'}$ for $s\bar q$. Notice that $\hat{g}$
also couples isospin $0$ $Q\bar{Q}$ states. Matrix elements of
spin and color operators are given in Appendix~\ref{app:ins}. The values
of $g$ and $g^{\prime}$ can be found in Ref~\cite{Yuan:2012wz},
notice that a factor $6$ is included in $g$ and $g^{\prime}$ in~\cite{Yuan:2012wz}.

\section{Numerical results}
\label{sec:result}

As stated in Sec~\ref{sec:frame},
parameters in the present manuscript are the constituent
masses of the light and strange quarks $m$ and $m_{s}$, the unharmonic part of
the confinement potential $V_{0}$,
the oscillator parameter $\omega$, and the strength
constants in the employed three kinds of hyperfine interactions.
We just take the parameters for hyperfine interactions to be the ones
given in the literatures listed in Sec~\ref{sec:frame}, and the
constituent masses of quarks and $V_{0}$ to be the empirical ones
given in~\cite{Yuan:2012wz}, and the oscillator parameter $\omega$
is then $\sim196$ MeV. All the values for parameters
are listed in Table~\ref{para}.

The procedure to obtain numerical results in our model can be
via three steps: at first, one has to calculate the eigenvalues
of the orbital Hamiltonian $H_{o}$, which can be
obtained by Eqs.~(\ref{eng}) and~(\ref{mcorrt}); secondly,
one should calculate the matrix elements of the
perturbative hyperfine interactions,  as we can see in Table~\ref{con},
matrices for the spin $1/2$, $3/2$ and $5/2$ configurations
are with 4, 4 and 1 dimensions, respectively; finally,
energies of the studied states should be produced
by diagonalization of the obtained matrices in the second step.

In a given model, all of the studied five-quark configurations should
have a same energy if we don't consider the hyperfine interactions
between quarks and difference between constituent masses of
light and strange quarks. Since the unharmonic parameter $V_{0}$
is taken to be different in the three hyperfine interaction model~\cite{Yuan:2012wz},
the obtained numerical values $E_{0}\equiv\sum_{i=1}^{5}m_{i}+E_{o}$
for the five-quark configurations with $q\bar{q}$
in the OGE, GBE and INS models are
$2196$, $1891$ and $2171$ MeV, respectively, and those for
the configurations with $s\bar{s}$ are $2436$, $2131$ and $2411$ MeV
in the three models, respectively.

In the calculations of elements of the hyperfine interactions matrices,
we label the configurations listed in Table~\ref{con} as follow
\begin{eqnarray}
1&:&s^{3}q([4]_{X}[211]_{C}[31]_{FS}[31]_{F}[22]_{S})\otimes \bar{q}\,,\nonumber\\
2&:&s^{3}q([4]_{X}[211]_{C}[31]_{FS}[31]_{F}[31]_{S})\otimes \bar{q}\,,\nonumber\\
3&:&s^{3}q([4]_{X}[211]_{C}[31]_{FS}[4]_{F}[31]_{S})\otimes \bar{q}\,,\nonumber\\
4&:&s^{4}([4]_{X}[211]_{C}[31]_{FS}[4]_{F}[31]_{S})\otimes \bar{s}\,,
\label{num12}
\end{eqnarray}
for the configurations with total spin $1/2$, and
\begin{eqnarray}
1&:&s^{3}q([4]_{X}[211]_{C}[31]_{FS}[31]_{F}[31]_{S})\otimes \bar{q}\,,\nonumber\\
2&:&s^{3}q([4]_{X}[211]_{C}[31]_{FS}[31]_{F}[4]_{S})\otimes \bar{q}\,,\nonumber\\
3&:&s^{3}q([4]_{X}[211]_{C}[31]_{FS}[4]_{F}[31]_{S})\otimes \bar{q}\,,\nonumber\\
4&:&s^{4}([4]_{X}[211]_{C}[31]_{FS}[4]_{F}[31]_{S})\otimes \bar{s}\,,
\label{num32}
\end{eqnarray}
for the configurations with total spin $3/2$. We give the numerical results
for the energies of these configurations in the next two subsections.

\subsection{Numerical results without corrections of configuration mixing}

Explicit calculations of the matrix elements of hyperfine interactions
between quarks in the three models lead to the following matrices
for the spin $1/2$ and $3/2$ five-quark configurations
{\footnotesize
\begin{eqnarray}
\mathcal{E}_{1/2}^{OGE}=
\pmatrix{ 2235.0  &  -139.6 & 10.8 & 0.0 \cr
          -149.6  &  2365.4 & -25.6 & 0.0\cr
            10.8  &  -25.6  &  2373.7 & 0.0\cr
            0.0   & 0.0     &   0.0   & 2654.7\cr} \,,
\mathcal{E}_{3/2}^{OGE}=
\pmatrix{ 2223.4  &  328.9 & 6.6 & 0.0 \cr
          328.9  &  2095.0 & -68.0 & 0.0\cr
            6.6  &  -68.0  &  2333.7 & 0.0\cr
            0.0   & 0.0     &   0.0   & 2517.1\cr} \,,
\nonumber\\
\mathcal{E}_{1/2}^{GBE}=
\pmatrix{ 1833.6  &  0.0  & 0.0 & 0.0 \cr
           0.0  &  1896.6 & -16.2 & 0.0\cr
            0.0  &  -16.2  &  2010.0 & 0.0\cr
            0.0   & 0.0     &   0.0   & 2161.6\cr} \,,
\mathcal{E}_{3/2}^{GBE}=
\pmatrix{ 1896.6  &  0.0  & -16.2 & 0.0 \cr
           0.0  &  1990.2 & 0.0 & 0.0\cr
           -16.2  &  0.0  &  2010.0 & 0.0\cr
            0.0   & 0.0     &   0.0   & 2161.6\cr} \,,
 \nonumber\\
\mathcal{E}_{1/2}^{INS}=
\pmatrix{ 1928.0  &  -121.5 & -30.4 & -33.3 \cr
          -121.5  &   1908.8 & 30.4 & 33.3  \cr
           -30.4  &   30.4   & 2230.3 & 47.1 \cr
           -33.3  &   33.3     &  47.1   & 2411.0\cr} \,,
\mathcal{E}_{3/2}^{INS}=
\pmatrix{ 2052.0  &  -113.3 & -60.7 & -66.7 \cr
          -113.3  &   2250.0 & 191.9 & 210.8  \cr
           -60.7  &   191.9   & 2159.0 & 188.5 \cr
           -66.7  &   210.8     &  188.5   & 2411.0\cr} \,.
\label{engm}
\end{eqnarray}}
The results are in unit of MeV. Notice that values of $E_{0}$
are already included in the diagonal terms of the matrices.
For the only configuration with spin $5/2$, the numerical results
in OGE, GBE and INS models are $2492$, $1990$ and $1987$ MeV,
respectively.

As we can see in Eq.~(\ref{engm}), in the OGE model,
it's very interesting that the diagonal matrix element with lowest
energy in $\mathcal{E}_{1/2}^{OGE}$
is about $140$ MeV higher than that in $\mathcal{E}_{3/2}^{OGE}$.
As we know, the diagonal matrix elements $\mathcal{E}_{S}^{hyp}(i,i)$
in the above matrices just represent the energies
of the $ith$ five-quark configuration as labeled by Eqs.~(\ref{num12}) and~(\ref{num32}),
so we can conclude that the lowest energy state in OGE model is with spin $3/2$,
if mixing of the configurations with same spin are not taken into account.
It's because of that the contributions from $Q\bar{Q}$ hyperfine interactions
to spin $1/2$ states are larger than those to $3/2$ states.
This is the most important difference between the OGE and the
other two models. In addition, as we can see in~(\ref{engm}),
some of the non-diagonal terms $M_{S}^{hyp}(i,j)$ are very large, which
should lead to strong mixing between different configurations.
While note that the configuration with $s\bar{s}$ pair doesn't
mix with the other three configurations in OGE model.

In the GBE model, there are only two nonzero non-diagonal matrix elements
in both of $\mathcal{E}_{1/2}^{GBE}$ and $\mathcal{E}_{3/2}^{GBE}$.
These nonzero elements are caused by the non-vanishing matrix elements
of the flavor dependent operator $C_{1,2}$ between flavor wave functions
$[31]_{F}$ and $[4]_{F}$.
But the nonzero $\mathcal{E}_{S}^{GBE}(i,j)$ are much smaller than the diagonal
matrix elements $\mathcal{E}_{S}^{GBE}(i,i)$, so the mixing of the configurations with same spin
should be very weak in GBE model. One may notice that an obvious
difference between OGE and GBE models is that $Q\bar{Q}$ interactions
are assumed to be included in meson-exchange between quarks in GBE model,
but in OGE model, $Q\bar{Q}$ interactions which lead to large non-diagonal
matrix elements are independent with hyperfine interactions between quark pairs.

In the INS model, the most significant feature is that the five-quark configurations
with $s\bar{s}$ can mix with those with light $q\bar{q}$, as we can see in
Eq.~(\ref{engm}), all the non-diagonal matrix elements $\mathcal{E}_{S}^{INS}(i,4)$
with $i=1,2,3$ are not $0$, this is caused by the nonzero matrix elements
$\langle q\bar{q}|\hat{g}_{i,5}|s\bar{s}\rangle$. Moreover, the matrix element
$\langle u\bar{u}|\hat{g}_{i,5}|d\bar{d}\rangle$ is also not $0$. In the present
manuscript, the studied five-quark configurations are with quantum numbers
of $\Omega^{*}$, whose isospin should be $0$, consequently,
in the case of $Q\bar{Q}=q\bar{q}$, wave function should be
the isocalar one $1/\sqrt{2}(|sssu\bar{u}\rangle+|sssd\bar{d}\rangle)$. Therefore,
the $Q\bar{Q}$ interaction between $\bar{u}$ and $\bar{d}$ components
should also contribute to the total energy.

\subsection{Numerical results with configurations mixings}

Diagonalization of matrices~(\ref{engm}) leads to the energies
for the studied five-quark configurations as shown in
Table~\ref{mass} compared to what
obtained by Helminen and Riska~\cite{Helminen}
using $SU(3)$ symmetric version of GBE model,
and the wave functions for spin $1/2$ and $3/2$
states correspond to the energies in Table~\ref{mass}
are of course linear combinations of the configurations
listed in Table~\ref{con},
and corresponding coefficients for the
combinations are given in Table~\ref{coe1} and~\ref{coe3}.

As we can see in Table~\ref{mass}, in OGE model, most of the
obtained energies after diagonalization of hyperfine interaction matrices
are lower than the diagonal matrix elements in Eq.~(\ref{engm}),
it means that mixing of the configurations decreases the
energies in OGE model. While in any case, spin of the state with lowest
energy is still $3/2$, which is about $328$ MeV lower than the
lowest spin $1/2$ state. Just as we can see in Eq.~(\ref{engm}),
absolute values of the non-diagonal matrix elements $\mathcal{E}_{3/2}^{OGE}(1,2)$
and $\mathcal{E}_{3/2}^{OGE}(2,1)$ are more than $2$
times of those of $\mathcal{E}_{1/2}^{OGE}(1,2)$
and $\mathcal{E}_{1/2}^{OGE}(2,1)$, and it's the same case
for $\mathcal{E}_{S}^{OGE}(2,3)$
and $\mathcal{E}_{S}^{OGE}(3,2)$, so mixing
of the spin $3/2$ states is much stronger than that
of spin $1/2$ states as shown in Table~\ref{coe1} and~\ref{coe3},
and energies of $3/2$ states are decreased
much more than $1/2$ states after diagonalization of matrices~(\ref{engm}).

In GBE model, the results don't change very much after diagonalization of Eq.~(\ref{engm}).
Because most of the non-diagonal matrix elements are $0$, and the only two nonzero
elements in $\mathcal{E}_{1/2}^{GBE}$ and $\mathcal{E}_{3/2}^{GBE}$ are very small,
so mixing of the configurations in GBE model is very weak, for instance, for the spin $1/2$
states, the only nonzero mixing angle which is for configurations 2 and 3 is about $0.14$.
Here we have compared our numerical results to what obtained by GBE model in flavor $SU(3)$
symmetric case~\cite{Helminen}. As we can see in Table~\ref{mass}, energies of the states
with $q\bar{q}$ in our model are around $100$ MeV lower than those obtained by Helminen and
Riska, and $s\bar{s}$ are around $220$ MeV lower. The reason is that on the one hand
the oscillator parameter is taken to be $\omega_{HR}=228$ MeV for all the five-quark configurations
in Helminen and Riska's model, while in the present manuscript, because
there are at least $3$ strange quarks in the
studied configurations, we employ $\omega=\sqrt{m/m_{s}}\omega_{HR}\simeq196$ MeV, and
take into account a correction~(\ref{mcorrt}) caused by lighter constituent mass of the
light quark, which should contribute about $83$ MeV to energies of all the studied configurations,
on the other hand, considering the corrections of $SU(3)$ symmetry breaking, the
strength for a $K$ (or $s\bar{s}$) exchange between one light and one strange (or two strange) quarks
should be smaller than strength for $\pi$ exchange between two light quarks,
but in Ref.~\cite{Helminen}, the strengths for $\pi$, $K$ and $\eta$
exchanges are taken to be a same one. To conclude, the model we employed here
is very close to the one used in Ref.~\cite{Helminen}, but there are two differences:
first, we have considered all the corrections of $SU(3)$ symmetry breaking which
have not been taken into account in Ref.~\cite{Helminen}, including the
corrections to the harmonic oscillator parameter, and the different 
strengths for different mesons exchange; secondly, in Ref.~\cite{Helminen}, mixing 
between configurations with same quantum number is neglected, here we have given
the corrections of configurations mixing explicitly, although its effects are very
small.

In INS model, mixing of the configurations with same spin parity quantum number
is very strong, as shown in Table~\ref{coe1} and~\ref{coe3}.
As we can see in Table~\ref{mass}, energy of the spin $1/2$ state
with lowest energy is about $131$ MeV lower than the one without configuration mixing
corrections given in Eq.~(\ref{engm}), that is caused by strong mixing between first
and second configurations labeled by Eq.~(\ref{num12}) and~(\ref{num32}).
On the other hand, the $Q\bar{Q}$ INS interactions lead to mixing between
configurations with $q\bar{q}$ and $s\bar{s}$ pairs, as shown in Table~\ref{coe3},
this kind of mixing between the spin $3/2$ configurations is very strong.

Note that it is very difficult to compare our numerical results to
the existing experimental data, because the spin parity quantum
numbers of the experimentally discovered $\Omega^{*}$ have not been
clarified. Here we just tentatively compare the present results to
what obtained in different three-quark models, which are listed in
Table~\ref{omega}. As we can see, the lowest energies in the present
three five-quark models are $\sim$ 200 MeV lower than most of those
listed in Table~\ref{omega}, just as we have expected. Accordingly,
five-quark components may be the more preferable ones in
$\Omega^{*}$ than traditional three-quark excitations, and this can
be examined by measurements of $\Omega^*\bar{\Omega}$ production in
$\Psi^{\prime}$ decay performed at BESIII~\cite{Asner:2008nq}.

Although the predictions listed in Table~\ref{omega} are from
different models from the ones we use in this manuscript, 
at least, we still can compare our numerical results to the ones
from Refs.~\cite{CIK81,GR96b}, since two of the employed models
in the present manuscript are just developed based on the
ones used in~\cite{CIK81,GR96b}, from the $Q^{3}$ three-quark case
to $Q^{4}\bar{Q}$ case. While on may notice that the obtained numerical
results in our manuscript should strongly depend on the hyperfine
interaction parameters, especially the OGE and INS models, since
the mixing of the considered configurations are very strong. Therefore, 
although values for the parameters we have used are either from
literatures or some physical bounds, which should be fairly reliable,
here we vary values of the hyperfine interaction parameters
listed in Table~\ref{para} by $0$ to $\pm20\%$
to see the dependence of our results on the parameters.
Resulting numerical results are given in Table~\ref{mass2}.
As we can see in the table, numerical results in OGE and
INS models do change a lot, but our main conclusion don't change, namely,
lowest energy of five-quark configuration is lower than
that of three-quark configuration, and the lowest energy state
in OGE model is with spin $3/2$.

\section{Conclusion}

\label{sec:end}

To conclude the present manuscript, here we calculate
the energies of $9$ low-lying $s^{3}Q\bar{Q}$
configurations with quantum numbers of $\Omega^{*}$. One gluon exchange,
Goldstone Boson exchange and instanton-induced interactions are taken as
hyperfine interactions between quarks, respectively. In addition, corrections
of flavor $SU(3)$ symmetry breaking are also taken into account.

The numerical results show that the OGE and INS hyperfine interactions
lead to strong mixing of the configurations with same spin parity, and
decrease the energies of the several studied states a lot. In OGE model, it's very
interesting that the lowest energy state is with spin $3/2$ but not $1/2$,
this is different from results obtained by GBE and INS models. It's because 
of that on the one hand the contributions from the diagonal matrix elements 
of hyperfine interactions caused by OGE to the energies of spin $1/2$ states
are larger than those to the energies of spin $3/2$ states, and on the other hand,
the nondiagonal matrix elements of OGE hyperfine interactions 
which lead to configurations mixing decrease
energy of the first spin $3/2$ state much more than energy of the first spin $1/2$ state, 
while the other two hyperfine interactions don't have these features. 
In GBE model, 
large discrepancies of our results from what obtained in Ref.~\cite{Helminen}
indicate that the $SU(3)$ breaking corrections are significant. In INS model,
$Q\bar{Q}$ hyperfine interactions lead to strong mixing between five-quark
configurations with light $q\bar{q}$ and strange $s\bar{s}$, this is the most
significant difference between INS and the other two models.

It is not convenient for us to compare our results to experimental
data, because the data is very poor. While comparing to predictions
of traditional three-quark models, the state with lowest energy in
our model is $\sim200$ MeV lower, it indicates that the energy cost
to excite ground state of $\Omega$ hyperon to a five-quark state is
less than that to an orbital excitation. Accordingly, five-quark
components may be more preferable ones in $\Omega^{*}$. If this
conclusion is correct, namely, the lowest $\Omega^{*}$ is lying at
$\sim1800$ MeV, then it can be found in measurements of
$\Psi^{\prime}$ decays performed at BESIII. Moreover, if
$\Omega^{*}$ with lowest energy is experimentally found at
$\sim1800$ MeV, spin parity quantum number of this $\Omega$
excitation could be an evidence to indicate whether the OGE model or
the other two hyperfine interaction models is more appropriate in
studies of $\Omega$ excitations.

On the other hand, here we have only considered the configurations with negative
parity, work on energies of the $s^{3}Q\bar{Q}$ configurations with
positive parity is in progress.

%\newpage
\section*{Acknowledgements}
This work was supported by the National Natural Science Foundation
of China (Grant Nos. 11205164, 10905059, 10925526, 11035006 and 11147197), DFG
and NSFC (CRC 110), the Chinese Academy of Sciences under Project
No.KJCX2-EW-N01 and the Ministry of Science and Technology of China
(2009CB825200).

%%%%%%%%%%%%%%%%%%%%%%%%%%%%%%%%%%%%%%%%%%%%%%%%%%%%%%%%%%%%%%%%%%%%%%%%%%%%%%%%%%%%%%%%%%%%%%%%%%%%%%%%%%%
%                                 References                                                              %
%%%%%%%%%%%%%%%%%%%%%%%%%%%%%%%%%%%%%%%%%%%%%%%%%%%%%%%%%%%%%%%%%%%%%%%%%%%%%%%%%%%%%%%%%%%%%%%%%%%%%%%%%%%

%%%%%%%%%%%%%%%%%%%%%%%%%%%%%%%%%%%%%%%%%%%%%%%%%%%%%%%%%%%%%%%%%%%%%%%%%%%%%%%%%%%%%%%%%%%%%%%%%%%%%%%%%%%%%%%%%%%%%%%%%

%%%%%%%%%%%%%%%%%%%%%%%%%%%%%%%%%%%%%%%%%%%%%%%%%%%%%%%%%%%%%%%%%%%%%%%%%%%%%%%%%%%%%%%%%%%%%%%%%%%%%%%%%%%%%%%%%%%%%%%%%
%                                           Appendix                                                                    %
%%%%%%%%%%%%%%%%%%%%%%%%%%%%%%%%%%%%%%%%%%%%%%%%%%%%%%%%%%%%%%%%%%%%%%%%%%%%%%%%%%%%%%%%%%%%%%%%%%%%%%%%%%%%%%%%%%%%%%%%%

\newpage

\begin{appendix}

\section{Matrix elements of the operators in OGE model}

\label{app:oge}

Matrix elements of the color operators $\vec\lambda_{1}^{C}\cdot\vec{\lambda}_{2}^{C}$
and $-\vec\lambda_{4}^{C}\cdot\vec\lambda_{5}^{C*}$ between
three color states are
\begin{eqnarray}
&\langle[211]_{C1}\otimes C_{\bar{Q}}|\vec\lambda_{1}^{C}
\cdot\vec{\lambda}_{2}^{C}|[211]_{C1}\otimes C_{\bar{Q}}\rangle=4/3\,,\nonumber\\
&\langle[211]_{C2}\otimes C_{\bar{Q}}|\vec\lambda_{1}^{C}
\cdot\vec{\lambda}_{2}^{C}|[211]_{C2}\otimes C_{\bar{Q}}\rangle=-8/3\,,\nonumber\\
&\langle[211]_{C3}\otimes C_{\bar{Q}}|\vec\lambda_{1}^{C}
\cdot\vec{\lambda}_{2}^{C}|[211]_{C3}\otimes C_{\bar{Q}}\rangle=-8/3\,,\nonumber\\
&\langle[211]_{C1}\otimes C_{\bar{Q}}|-\vec\lambda_{4}^{C}
\cdot\vec{\lambda}_{5}^{C*}|[211]_{C1}\otimes C_{\bar{Q}}\rangle=2/3\,,\nonumber\\
&\langle[211]_{C2}\otimes C_{\bar{Q}}|-\vec\lambda_{4}^{C}
\cdot\vec{\lambda}_{5}^{C*}|[211]_{C2}\otimes C_{\bar{Q}}\rangle=2/3\,,\nonumber\\
&\langle[211]_{C3}\otimes C_{\bar{Q}}|-\vec\lambda_{4}^{C}
\cdot\vec{\lambda}_{5}^{C*}|[211]_{C3}\otimes C_{\bar{Q}}\rangle=-16/3\,.
\end{eqnarray}
Matrix elements of the spin operators $\vec\sigma_{1}\cdot\vec\sigma_{2}$ which are independent
with the total angular momentum of the five quark configurations are
\begin{eqnarray}
\langle S_{5}([22]_{S1}\otimes S_{\bar{Q}})|\vec\sigma_{1}\cdot\vec\sigma_{2}|S_{5}([22]_{S1}\otimes S_{\bar{Q}})\rangle=1\,,\nonumber\\
\langle S_{5}([22]_{S2}\otimes S_{\bar{Q}})|\vec\sigma_{1}\cdot\vec\sigma_{2}|S_{5}([22]_{S2}\otimes S_{\bar{Q}})\rangle=-3\,,\nonumber\\
\langle S_{5}([31]_{S1}\otimes S_{\bar{Q}})|\vec\sigma_{1}\cdot\vec\sigma_{2}|S_{5}([31]_{S1}\otimes S_{\bar{Q}})\rangle=1\,,\nonumber\\
\langle S_{5}([31]_{S2}\otimes S_{\bar{Q}})|\vec\sigma_{1}\cdot\vec\sigma_{2}|S_{5}([31]_{S2}\otimes S_{\bar{Q}})\rangle=1\,,\nonumber\\
\langle S_{5}([31]_{S3}\otimes S_{\bar{Q}})|\vec\sigma_{1}\cdot\vec\sigma_{2}|S_{5}([31]_{S3}\otimes S_{\bar{Q}})\rangle=-3\,,\nonumber\\
\langle S_{5}([4]_{S}\otimes S_{\bar{Q}})|\vec\sigma_{1}\cdot\vec\sigma_{2}|S_{5}([4]_{S}\otimes S_{\bar{Q}})\rangle=1\,,
\end{eqnarray}
and those of $\vec{\sigma}_{4}\cdot\vec{\sigma}_{5}$ are
\begin{eqnarray}
\langle 1/2([31]_{S1}\otimes S_{\bar{Q}})|\vec\sigma_{4}\cdot\vec\sigma_{5}|1/2([31]_{S1}\otimes S_{\bar{Q}})\rangle=1\,,\nonumber\\
\langle 1/2([31]_{S2}\otimes S_{\bar{Q}})|\vec\sigma_{4}\cdot\vec\sigma_{5}|1/2([31]_{S2}\otimes S_{\bar{Q}})\rangle=-2\,,\nonumber\\
\langle 1/2([31]_{S3}\otimes S_{\bar{Q}})|\vec\sigma_{4}\cdot\vec\sigma_{5}|1/2([31]_{S3}\otimes S_{\bar{Q}})\rangle=-2\,,\nonumber\\
\langle 1/2([22]_{S1}\otimes S_{\bar{Q}})|\vec\sigma_{4}\cdot\vec\sigma_{5}|1/2([31]_{S2}\otimes S_{\bar{Q}})\rangle=\sqrt{3}\,,\nonumber\\
\langle 1/2([22]_{S2}\otimes S_{\bar{Q}})|\vec\sigma_{4}\cdot\vec\sigma_{5}|1/2([31]_{S3}\otimes S_{\bar{Q}})\rangle=\sqrt{3}\,,\nonumber\\
\langle 3/2([31]_{S1}\otimes S_{\bar{Q}})|\vec\sigma_{4}\cdot\vec\sigma_{5}|3/2([31]_{S1}\otimes S_{\bar{Q}})\rangle=-1/2\,,\nonumber\\
\langle 3/2([31]_{S2}\otimes S_{\bar{Q}})|\vec\sigma_{4}\cdot\vec\sigma_{5}|3/2([31]_{S2}\otimes S_{\bar{Q}})\rangle=1\,,\nonumber\\
\langle 3/2([31]_{S3}\otimes S_{\bar{Q}})|\vec\sigma_{4}\cdot\vec\sigma_{5}|3/2([31]_{S3}\otimes S_{\bar{Q}})\rangle=1\,,\nonumber\\
\langle 3/2([4]_{S}\otimes S_{\bar{Q}})|\vec\sigma_{4}\cdot\vec\sigma_{5}|3/2([4]_{S}\otimes S_{\bar{Q}})\rangle=-3/2\,,\nonumber\\
\langle 3/2([31]_{S1}\otimes S_{\bar{Q}})|\vec\sigma_{4}\cdot\vec\sigma_{5}|3/2([4]_{S}\otimes S_{\bar{Q}})\rangle=\sqrt{15}/2\,,\nonumber\\
\langle 5/2([4]_{S}\otimes S_{\bar{Q}})|\vec\sigma_{4}\cdot\vec\sigma_{5}|5/2([4]_{S}\otimes S_{\bar{Q}})\rangle=1\,.
\end{eqnarray}
Finally, matrix elements of the flavor dependent chromomagnetic strenth operators $C_{1,2}$ and $C_{4,5}$ are as follow
\begin{eqnarray}
&\langle s^{3}q([31]_{F1})\otimes\bar{q}|C_{1,2}| s^{3}q([31]_{F1})\otimes\bar{q}\rangle=\frac{1}{6}(5C_{ss}+C_{qs})\,,\nonumber\\
&\langle s^{3}q([31]_{F2})\otimes\bar{q}|C_{1,2}| s^{3}q([31]_{F2})\otimes\bar{q}\rangle=\frac{1}{3}(2C_{ss}+C_{qs})\,,\nonumber\\
&\langle s^{3}q([31]_{F3})\otimes\bar{q}|C_{1,2}| s^{3}q([31]_{F3})\otimes\bar{q}\rangle=C_{qs},\nonumber\\
&\langle s^{3}q([31]_{F1})\otimes\bar{q}|C_{1,2}| s^{3}q([31]_{F2})\otimes\bar{q}\rangle
=\frac{1}{3\sqrt{2}}(C_{qs}-C_{ss})\,,\nonumber\\
&\langle s^{3}q([31]_{F2})\otimes\bar{q}|C_{1,2}| s^{3}q([31]_{F1})\otimes\bar{q}\rangle=
\frac{1}{3\sqrt{2}}(C_{qs}-C_{ss})\,,\nonumber\\
&\langle s^{3}q([4]_{F})\otimes\bar{q}|C_{1,2}|s^{3}q([4]_{F})\otimes\bar{q}\rangle=\frac{1}{2}(C_{qs}+C_{ss})\,,\nonumber\\
&\langle s^{4}([4]_{F})\otimes\bar{s}|C_{1,2}|s^{4}([4]_{F})\otimes\bar{s}\rangle=C_{ss}\,,\nonumber\\
&\langle s^{3}q([31]_{F1})\otimes\bar{q}|C_{1,2}|s^{3}q([4]_{F})\otimes\bar{q}\rangle=
\frac{\sqrt{3}}{6}(C_{qs}-C_{ss})\,,\nonumber\\
&\langle s^{3}q([31]_{F2})\otimes\bar{q}|C_{1,2}|s^{3}q([4]_{F})\otimes\bar{q}\rangle=
\frac{\sqrt{6}}{6}(C_{qs}-C_{ss})\,,\nonumber\\
&\langle s^{3}q([31]_{F1})\otimes\bar{q}|C_{4,5}| s^{3}q([31]_{F1})\otimes\bar{q}\rangle=\frac{1}{4}(3C_{\bar{q}q}+C_{\bar{q}s})\,,\nonumber\\
&\langle s^{3}q([31]_{F2})\otimes\bar{q}|C_{4,5}| s^{3}q([31]_{F2})\otimes\bar{q}\rangle=C_{\bar{q}s}\,,\nonumber\\
&\langle s^{3}q([31]_{F3})\otimes\bar{q}|C_{4,5}| s^{3}q([31]_{F3})\otimes\bar{q}\rangle=C_{\bar{q}s},\nonumber\\
&\langle s^{3}q([4]_{F})\otimes\bar{q}|C_{4,5}| s^{3}q([4]_{F})\otimes\bar{q}\rangle=\frac{1}{4}(C_{\bar{q}q}+3C_{\bar{q}s})\,,\nonumber\\
&\langle s^{4}([4]_{F})\otimes\bar{s}|C_{4,5}| s^{4}([4]_{F})\otimes\bar{s}\rangle=C_{\bar{s}s}\,.
\end{eqnarray}
Notice that here we have only listed the nonzero matrix elements.

\section{Matrix elements of the flavor dependent operator in GBE model}
\label{app:gbe}

Matrix elements of the flavor dependent operator $C_{1,2}^{M}\vec{\lambda}_{1}^{F}\cdot\vec{\lambda}_{2}^{F}$
are as follow
\begin{eqnarray}
&\langle s^{3}q([31]_{F1})\otimes\bar{q}|C_{1,2}| s^{3}q([31]_{F1})\otimes\bar{q}\rangle
=\frac{1}{9}(10C^{s\bar{s}}+2C^{K})\,,\nonumber\\
&\langle s^{3}q([31]_{F2})\otimes\bar{q}|C_{1,2}| s^{3}q([31]_{F2})\otimes\bar{q}\rangle
=\frac{1}{9}(8C^{s\bar{s}}+4C^{K})\,,\nonumber\\
&\langle s^{3}q([31]_{F3})\otimes\bar{q}|C_{1,2}| s^{3}q([31]_{F3})\otimes\bar{q}\rangle
=-\frac{8}{3}C^{K},\nonumber\\
&\langle s^{3}q([31]_{F1})\otimes\bar{q}|C_{1,2}| s^{3}q([31]_{F2})\otimes\bar{q}\rangle
=\frac{2\sqrt{2}}{9}(C^{K}-C^{s\bar{s}})\,,\nonumber\\
&\langle s^{3}q([31]_{F2})\otimes\bar{q}|C_{1,2}| s^{3}q([31]_{F1})\otimes\bar{q}\rangle=
\frac{2\sqrt{2}}{9}(C^{K}-C^{s\bar{s}})\,,\nonumber\\
&\langle s^{3}q([4]_{F})\otimes\bar{q}|C_{1,2}|s^{3}q([4]_{F})\otimes\bar{q}\rangle
=\frac{2}{3}(C^{s\bar{s}}+C_{K})\,,\nonumber\\
&\langle s^{4}([4]_{F})\otimes\bar{s}|C_{1,2}|s^{4}([4]_{F})\otimes\bar{s}\rangle=
\frac{4}{3}C^{s\bar{s}}\,,\nonumber\\
&\langle s^{3}q([31]_{F1})\otimes\bar{q}|C_{1,2}|s^{3}q([4]_{F})\otimes\bar{q}\rangle=
\frac{2}{3\sqrt{3}}(C^{K}-C^{s\bar{s}})\,,\nonumber\\
&\langle s^{3}q([31]_{F2})\otimes\bar{q}|C_{1,2}|s^{3}q([4]_{F})\otimes\bar{q}\rangle=
\frac{4}{3\sqrt{6}}(C^{K}-C^{s\bar{s}})\,.
\end{eqnarray}

\section{Matrix elements of the operators in the INS model}
\label{app:ins}

Matrix elements of the color operators $P_{1,2}^{C,6}$,
$P_{1,2}^{C,\bar{3}}$, $P_{4,5}^{C,8}$ and $P_{4,5}^{C,1}$
between three color states are
\begin{eqnarray}
&\langle[211]_{C1}\otimes C_{\bar{Q}}|P_{1,2}^{C,6}|[211]_{C1}\otimes C_{\bar{Q}}\rangle=1\,,\nonumber\\
&\langle[211]_{C2}\otimes C_{\bar{Q}}|P_{1,2}^{C,6}|[211]_{C2}\otimes C_{\bar{Q}}\rangle=0\,,\nonumber\\
&\langle[211]_{C3}\otimes C_{\bar{Q}}|P_{1,2}^{C,6}|[211]_{C3}\otimes C_{\bar{Q}}\rangle=0\,,\nonumber\\
&\langle[211]_{C1}\otimes C_{\bar{Q}}|P_{1,2}^{C,\bar{3}}|[211]_{C1}\otimes C_{\bar{Q}}\rangle=0\,,\nonumber\\
&\langle[211]_{C2}\otimes C_{\bar{Q}}|P_{1,2}^{C,\bar{3}}|[211]_{C2}\otimes C_{\bar{Q}}\rangle=1\,,\nonumber\\
&\langle[211]_{C3}\otimes C_{\bar{Q}}|P_{1,2}^{C,\bar{3}}|[211]_{C3}\otimes C_{\bar{Q}}\rangle=1\,,\nonumber\\
&\langle[211]_{C1}\otimes C_{\bar{Q}}|P_{4,5}^{C,8}|[211]_{C1}\otimes C_{\bar{Q}}\rangle=1\,,\nonumber\\
&\langle[211]_{C2}\otimes C_{\bar{Q}}|P_{4,5}^{C,8}|[211]_{C2}\otimes C_{\bar{Q}}\rangle=1\,,\nonumber\\
&\langle[211]_{C3}\otimes C_{\bar{Q}}|P_{4,5}^{C,8}|[211]_{C3}\otimes C_{\bar{Q}}\rangle=0\,,\nonumber\\
&\langle[211]_{C1}\otimes C_{\bar{Q}}|P_{4,5}^{C,1}|[211]_{C1}\otimes C_{\bar{Q}}\rangle=1\,,\nonumber\\
&\langle[211]_{C2}\otimes C_{\bar{Q}}|P_{4,5}^{C,1}|[211]_{C2}\otimes C_{\bar{Q}}\rangle=0\,,\nonumber\\
&\langle[211]_{C3}\otimes C_{\bar{Q}}|P_{4,5}^{C,1}|[211]_{C3}\otimes C_{\bar{Q}}\rangle=0\,.
\end{eqnarray}
Matrix elements of the spin operators $P_{1,2}^{S=0}$ and $P_{1,2}^{S=1}$ which are independent
with the total angular momentum of the five quark configurations are
\begin{eqnarray}
&\langle S_{5}([22]_{S1}\otimes S_{\bar{Q}})|P_{1,2}^{S=0}|S_{5}([22]_{S1}\otimes S_{\bar{Q}})\rangle=0\,,\nonumber\\
&\langle S_{5}([22]_{S2}\otimes S_{\bar{Q}})|P_{1,2}^{S=0}|S_{5}([22]_{S2}\otimes S_{\bar{Q}})\rangle=1\,,\nonumber\\
&\langle S_{5}([22]_{S1}\otimes S_{\bar{Q}})|P_{1,2}^{S=1}|S_{5}([22]_{S1}\otimes S_{\bar{Q}})\rangle=1\,,\nonumber\\
&\langle S_{5}([22]_{S2}\otimes S_{\bar{Q}})|P_{1,2}^{S=1}|S_{5}([22]_{S2}\otimes S_{\bar{Q}})\rangle=0\,,\nonumber\\
&\langle S_{5}([31]_{S1}\otimes S_{\bar{Q}})|P_{1,2}^{S=0}|S_{5}([31]_{S1}\otimes S_{\bar{Q}})\rangle=0\,,\nonumber\\
&\langle S_{5}([31]_{S2}\otimes S_{\bar{Q}})|P_{1,2}^{S=0}|S_{5}([31]_{S2}\otimes S_{\bar{Q}})\rangle=0\,,\nonumber\\
&\langle S_{5}([31]_{S3}\otimes S_{\bar{Q}})|P_{1,2}^{S=0}|S_{5}([31]_{S3}\otimes S_{\bar{Q}})\rangle=1\,,\nonumber\\
&\langle S_{5}([31]_{S1}\otimes S_{\bar{Q}})|P_{1,2}^{S=1}|S_{5}([31]_{S1}\otimes S_{\bar{Q}})\rangle=1\,,\nonumber\\
&\langle S_{5}([31]_{S2}\otimes S_{\bar{Q}})|P_{1,2}^{S=1}|S_{5}([31]_{S2}\otimes S_{\bar{Q}})\rangle=1\,,\nonumber\\
&\langle S_{5}([31]_{S3}\otimes S_{\bar{Q}})|P_{1,2}^{S=1}|S_{5}([31]_{S3}\otimes S_{\bar{Q}})\rangle=0\,,\nonumber\\
&\langle S_{5}([4]_{S}\otimes S_{\bar{Q}})|P_{1,2}^{S=0}|S_{5}([4]_{S}\otimes S_{\bar{Q}})\rangle=0\,,\nonumber\\
&\langle S_{5}([4]_{S}\otimes S_{\bar{Q}})|P_{1,2}^{S=1}|S_{5}([4]_{S}\otimes S_{\bar{Q}})\rangle=1\,,
\end{eqnarray}
and those of $P_{4,5}^{S=0}$ and $P_{4,5}^{S=1}$ are
\begin{eqnarray}
&\langle 1/2([22]_{S1}\otimes S_{\bar{Q}})|P_{4,5}^{S=0}|1/2([22]_{S1}\otimes S_{\bar{Q}})\rangle=1/4\,,\nonumber\\
&\langle 1/2([22]_{S2}\otimes S_{\bar{Q}})|P_{4,5}^{S=0}|1/2([22]_{S2}\otimes S_{\bar{Q}})\rangle=1/4\,,\nonumber\\
&\langle 1/2([22]_{S1}\otimes S_{\bar{Q}})|P_{4,5}^{S=1}|1/2([22]_{S1}\otimes S_{\bar{Q}})\rangle=3/4\,,\nonumber\\
&\langle 1/2([22]_{S2}\otimes S_{\bar{Q}})|P_{4,5}^{S=1}|1/2([22]_{S2}\otimes S_{\bar{Q}})\rangle=3/4\,,\nonumber\\
&\langle 1/2([31]_{S1}\otimes S_{\bar{Q}})|P_{4,5}^{S=0}|1/2([31]_{S1}\otimes S_{\bar{Q}})\rangle=0\,,\nonumber\\
&\langle 1/2([31]_{S2}\otimes S_{\bar{Q}})|P_{4,5}^{S=0}|1/2([31]_{S2}\otimes S_{\bar{Q}})\rangle=3/4\,,\nonumber\\
&\langle 1/2([31]_{S3}\otimes S_{\bar{Q}})|P_{4,5}^{S=0}|1/2([31]_{S3}\otimes S_{\bar{Q}})\rangle=3/4\,,\nonumber\\
&\langle 1/2([31]_{S1}\otimes S_{\bar{Q}})|P_{4,5}^{S=1}|1/2([31]_{S1}\otimes S_{\bar{Q}})\rangle=1\,,\nonumber\\
&\langle 1/2([31]_{S2}\otimes S_{\bar{Q}})|P_{4,5}^{S=1}|1/2([31]_{S2}\otimes S_{\bar{Q}})\rangle=1/4\,,\nonumber\\
&\langle 1/2([31]_{S3}\otimes S_{\bar{Q}})|P_{4,5}^{S=1}|1/2([31]_{S3}\otimes S_{\bar{Q}})\rangle=1/4\,,\nonumber\\
&\langle 1/2([22]_{S1}\otimes S_{\bar{Q}})|P_{4,5}^{S=0}|1/2([31]_{S2}\otimes S_{\bar{Q}})\rangle=-\sqrt{3}/4\,,\nonumber\\
&\langle 1/2([22]_{S2}\otimes S_{\bar{Q}})|P_{4,5}^{S=0}|1/2([31]_{S3}\otimes S_{\bar{Q}})\rangle=-\sqrt{3}/4\,,\nonumber\\
&\langle 1/2([22]_{S1}\otimes S_{\bar{Q}})|P_{4,5}^{S=1}|1/2([31]_{S2}\otimes S_{\bar{Q}})\rangle=\sqrt{3}/4\,,\nonumber\\
&\langle 1/2([22]_{S2}\otimes S_{\bar{Q}})|P_{4,5}^{S=1}|1/2([31]_{S3}\otimes S_{\bar{Q}})\rangle=\sqrt{3}/4\,,\nonumber\\
&\langle 3/2([31]_{S1}\otimes S_{\bar{Q}})|P_{4,5}^{S=0}|3/2([31]_{S1}\otimes S_{\bar{Q}})\rangle=3/8\,,\nonumber\\
&\langle 3/2([31]_{S2}\otimes S_{\bar{Q}})|P_{4,5}^{S=0}|3/2([31]_{S2}\otimes S_{\bar{Q}})\rangle=0\,,\nonumber\\
&\langle 3/2([31]_{S3}\otimes S_{\bar{Q}})|P_{4,5}^{S=0}|3/2([31]_{S3}\otimes S_{\bar{Q}})\rangle=0\,,\nonumber\\
&\langle 3/2([31]_{S1}\otimes S_{\bar{Q}})|P_{4,5}^{S=1}|3/2([31]_{S1}\otimes S_{\bar{Q}})\rangle=5/8\,,\nonumber\\
&\langle 3/2([31]_{S2}\otimes S_{\bar{Q}})|P_{4,5}^{S=1}|3/2([31]_{S2}\otimes S_{\bar{Q}})\rangle=1\,,\nonumber\\
&\langle 3/2([31]_{S3}\otimes S_{\bar{Q}})|P_{4,5}^{S=1}|3/2([31]_{S3}\otimes S_{\bar{Q}})\rangle=1\,,\nonumber\\
&\langle 3/2([4]_{S}\otimes S_{\bar{Q}})|P_{4,5}^{S=0}|3/2([4]_{S}\otimes S_{\bar{Q}})\rangle=5/8\,,\nonumber\\
&\langle 3/2([4]_{S}\otimes S_{\bar{Q}})|P_{4,5}^{S=1}|3/2([4]_{S}\otimes S_{\bar{Q}})\rangle=3/8\,,\nonumber\\
&\langle 3/2([31]_{S1}\otimes S_{\bar{Q}})|P_{4,5}^{S=0}|3/2([4]_{S}\otimes S_{\bar{Q}})\rangle=-\sqrt{15}/8\,,\nonumber\\
&\langle 3/2([31]_{S1}\otimes S_{\bar{Q}})|P_{4,5}^{S=1}|3/2([4]_{S}\otimes S_{\bar{Q}})\rangle=\sqrt{15}/8\,,\nonumber\\
&\langle 5/2([4]_{S}\otimes S_{\bar{Q}})|P_{4,5}^{S=0}|5/2([4]_{S}\otimes S_{\bar{Q}})\rangle=0\,,\nonumber\\
&\langle 5/2([4]_{S}\otimes S_{\bar{Q}})|P_{4,5}^{S=1}|5/2([4]_{S}\otimes S_{\bar{Q}})\rangle=1\,.
\end{eqnarray}

\end{appendix}

%%%%%%%%%%%%%%%%%%%%%%%%%%%%%%%%%%%%%%%%%%%%%%%%%%%%%%%%%%%%%%%%%%%%%%%%%%%%%%%%%%%%%%%%%%%%%%%%%%%%%%%%%%%%%%%%%%%%%%%%%%%%

% Tables
\newpage

%%%%%%%%%%%%%%%%%%%%%%%%%%%%%%%%%%%%%%%%%%%%%%%%%%%%%%%%%%%%%%%%%%%%%%%%%%%%%%%%%%%%%%%%%%%%%%%%%%%
%                            Table I                                                              %
%%%%%%%%%%%%%%%%%%%%%%%%%%%%%%%%%%%%%%%%%%%%%%%%%%%%%%%%%%%%%%%%%%%%%%%%%%%%%%%%%%%%%%%%%%%%%%%%%%%
\begin{table}[ht]
\caption{The studied $s^{3}Q\bar{Q}$ configurations with inner
quantum number $N_{r}=L=0$. Notice that Columns FS and CS
just denote the same configurations expressed in two different
languages.
\label{con}}
\vspace{0.3cm}
\renewcommand\tabcolsep{0.49cm}
\renewcommand{\arraystretch}{0.5}
\begin{tabular}{ccc}
\hline\hline
&&\\
FS     &     CS     & $J^{P}$ \\
&&\\
\hline
&&\\
$s^{3}q([4]_{X}[211]_{C}[31]_{FS}[31]_{F}[22]_{S})\otimes \bar{q}$ &
$s^{3}q([4]_{X}[31]_{F}[211]_{CS}[211]_{C}[22]_{S})\otimes \bar{q}$&  $\frac{1}{2}^{-}$\\

&&\\

$s^{3}q([4]_{X}[211]_{C}[31]_{FS}[31]_{F}[31]_{S})\otimes \bar{q}$ &
$s^{3}q([4]_{X}[31]_{F}[211]_{CS}[211]_{C}[31]_{S})\otimes \bar{q}$&  $\frac{1}{2}^{-}$\\

&  &  $\frac{3}{2}^{-}$\\
&&\\

$s^{3}q([4]_{X}[211]_{C}[31]_{FS}[31]_{F}[4]_{S})\otimes \bar{q}$ &
$s^{3}q([4]_{X}[31]_{F}[211]_{CS}[211]_{C}[4]_{S})\otimes \bar{q}$&  $\frac{3}{2}^{-}$\\

&  &  $\frac{5}{2}^{-}$\\
&&\\

$s^{3}q([4]_{X}[211]_{C}[31]_{FS}[4]_{F}[31]_{S})\otimes \bar{q}$ &
$s^{3}q([4]_{X}[4]_{F}[211]_{CS}[211]_{C}[31]_{S})\otimes \bar{q}$&  $\frac{1}{2}^{-}$\\

&  &  $\frac{3}{2}^{-}$\\
&&\\

$s^{4}([4]_{X}[211]_{C}[31]_{FS}[4]_{F}[31]_{S})\otimes \bar{s}$ &
$s^{4}([4]_{X}[4]_{F}[211]_{CS}[211]_{C}[31]_{S})\otimes \bar{s}$&  $\frac{1}{2}^{-}$\\

&  &  $\frac{3}{2}^{-}$\\
&&\\

\hline\hline
\end{tabular}
\end{table}
%%%%%%%%%%%%%%%%%%%%%%%%%%%%%%%%%%%%%%%%%%%%%%%%%%%%%%%%%%%%%%%%%%%%%%%%%%%%%%%%%%%%%%%%%%%%%%%%%%%
%%%%%%%%%%%%%%%%%%%%%%%%%%%%%%%%%%%%%%%%%%%%%%%%%%%%%%%%%%%%%%%%%%%%%%%%%%%%%%%%%%%%%%%%%%%%%%%%%%%

%%%%%%%%%%%%%%%%%%%%%%%%%%%%%%%%%%%%%%%%%%%%%%%%%%%%%%%%%%%%%%%%%%%%%%%%%%%%%%%%%%%%%%%%%%%%%%%%%%%%%%
%                               Table II                                                             %
%%%%%%%%%%%%%%%%%%%%%%%%%%%%%%%%%%%%%%%%%%%%%%%%%%%%%%%%%%%%%%%%%%%%%%%%%%%%%%%%%%%%%%%%%%%%%%%%%%%%%%
\begin{table}[t]
\caption{Values for the parameters in three kinds of
hyperfine interactions (in unit of MeV). \label{para}}
%\begin{ruledtabular}
\renewcommand\tabcolsep{0.27cm}
\renewcommand{\arraystretch}{1}
\begin{tabular}{ccccccccccc}
\hline \hline
OGE
& $m$   & 340 & $m_s$  & 460 &   $\omega$   & 196 & $V_0$  & -208\\

& $C_{qq}$  & 18.3 &  $C_{qs}$  & 11.2 & $C_{ss}$ & 6.8 & $C_{q \bar q}$ & 29.8\\

           & $C_{q \bar s}$ & 18.4  &  $C_{s \bar s}$ & 8.6 & & \\

GBE
& $m$ &  340 & $m_s$  &  460   &  $\omega$ & 196  & $V_0$  &-269  \\

&  $C^{\pi}$ & 21 &&
\\

INS
& $m$ & 340 & $m_s$ &  460 &  $\omega$  &  196  &  $V_0$  &  -213\\

& $g$  &  52.5  &  $g^{\prime}$  & 33.3 \\

\hline\hline
\end{tabular}
%\end{ruledtabular}
\end{table}
%%%%%%%%%%%%%%%%%%%%%%%%%%%%%%%%%%%%%%%%%%%%%%%%%%%%%%%%%%%%%%%%%%%%%%%%%%%%%%%%%%%%%%%%%%%%%%%%%%%%%%%%%%

%%%%%%%%%%%%%%%%%%%%%%%%%%%%%%%%%%%%%%%%%%%%%%%%%%%%%%%%%%%%%%%%%%%%%%%%%%%%%%%%%%%%%%%%%%%%%%%%%%%%%%%
%                              Table III                                                              %
%%%%%%%%%%%%%%%%%%%%%%%%%%%%%%%%%%%%%%%%%%%%%%%%%%%%%%%%%%%%%%%%%%%%%%%%%%%%%%%%%%%%%%%%%%%%%%%%%%%%%%%
\begin{table}[t]
\caption{Energies of the studied five-quark configurations
in the three models, compared to the results obtained in~\cite{Helminen}
listed in column HR. The results are in unit of MeV.
\label{mass}}
\renewcommand\tabcolsep{0.88cm}
\renewcommand{\arraystretch}{1}

\vspace{0.3cm}

\begin{tabular}{cccccc} \hline\hline

$J^P$ & OGE &  GBE  & INS  &   HR \\

\hline

${1 \over 2}^-$

& 2146   &   1834  & 1797  & 1917\\

& 2451   &   1894  & 2027  & 1973\\

& 2677   &   2012  & 2223  & 2141\\

&  2655  &   2162  & 2431  & 2381\\

${3 \over 2}^-$

& 1818   &  1894   &  1987 & 1973\\

& 2331   &  1990   & 2025  & 2085\\

& 2503   &  2012   & 2143  & 2141\\

& 2517  &   2162   & 2717  & 2381\\

${5 \over 2}^-$

& 2492 &   1990   & 1987  & 2085\\

\hline
\hline
\end{tabular}
\end{table}
%%%%%%%%%%%%%%%%%%%%%%%%%%%%%%%%%%%%%%%%%%%%%%%%%%%%%%%%%%%%%%%%%%%%%%%%%%%%%%%%%%%%%%%%%%%%%%%%%%%%%%%%%%%%%%%%

\newpage
%%%%%%%%%%%%%%%%%%%%%%%%%%%%%%%%%%%%%%%%%%%%%%%%%%%%%%%%%%%%%%%%%%%%%%%%%%%%%%%%%%%%%%%%%%%%%%%%%%%%%%%%%%%%%%%%
%                                   Table IV                                                                   %
%%%%%%%%%%%%%%%%%%%%%%%%%%%%%%%%%%%%%%%%%%%%%%%%%%%%%%%%%%%%%%%%%%%%%%%%%%%%%%%%%%%%%%%%%%%%%%%%%%%%%%%%%%%%%%%%

\begin{table}[t]
\caption{Coefficients for the mixings of the configurations
with spin $1/2$ in three models.
\label{coe1}}
\renewcommand\tabcolsep{0.88cm}
\renewcommand{\arraystretch}{1}

\vspace{0.3cm}

\begin{tabular}{cccccc} \hline\hline

HYP & 1 &  2  & 3  &   4 \\

\hline

OGE

& 0.843   &   0.538  & 0.009   & 0   \\

& -0.535   &   0.836  & 0.122  & 0  \\

& 0.058   &   -0.108  & 0.992  & 0  \\

& 0      &      0  &    0  & 1  \\

GBE

& 1   &  0   &  0     &   0     \\

& 0   &  0.990   & 0.138  & 0    \\

& 0    &  -0.138   & 0.990  & 0  \\

& 0  &   0   & 0  & 1      \\

INS

& 0.678   &  0.735   &  -0.004     &   -0.003     \\

& 0.720   &  -0.662   & 0.184  & 0.098    \\

& -0.109    &  0.105   & 0.949  & 0.276  \\

& -0.103  &   0.010   & 0.256  & 0.956      \\

\hline
\hline
\end{tabular}
\end{table}
%%%%%%%%%%%%%%%%%%%%%%%%%%%%%%%%%%%%%%%%%%%%%%%%%%%%%%%%%%%%%%%%%%%%%%%%%%%%%%%%%%%%%%%%%%%%%%%%%%%%%%%

%%%%%%%%%%%%%%%%%%%%%%%%%%%%%%%%%%%%%%%%%%%%%%%%%%%%%%%%%%%%%%%%%%%%%%%%%%%%%%%%%%%%%%%%%%%%%%%%%%%%%%%%
%                                  Table V                                                             %
%%%%%%%%%%%%%%%%%%%%%%%%%%%%%%%%%%%%%%%%%%%%%%%%%%%%%%%%%%%%%%%%%%%%%%%%%%%%%%%%%%%%%%%%%%%%%%%%%%%%%%%%

\begin{table}[ht]
\caption{Coefficients for the mixings of the configurations
with spin $3/2$ in three models.
\label{coe3}}
\renewcommand\tabcolsep{0.88cm}
\renewcommand{\arraystretch}{1}

\vspace{0.3cm}

\begin{tabular}{cccccc} \hline\hline

HYP & 1 &  2  & 3  &   4 \\

\hline

OGE

& -0.627   &   0.771  & 0.110   & 0   \\

& 0.243   &   0.060  & 0.968  & 0  \\

& 0.740   &   0.634  & -0.226  & 0  \\

& 0      &      0  &    0  & 1  \\

GBE

& 0.990   &  0   &  0.138     &   0     \\

& 0   &  1  & 0  & 0    \\

& -0.138    &  0   & 0.990  & 0  \\

& 0  &   0   & 0  & 1      \\

INS

& 0.646   &  0.612   &  -0.456     &   0.000     \\

& 0.647   &  -0.138   & 0.730  & -0.170    \\

& 0.352    &  -0.560   & -0.254  & 0.706  \\

& -0.201  &   0.541   & 0.440  & 0.688      \\

\hline
\hline
\end{tabular}
\end{table}
%%%%%%%%%%%%%%%%%%%%%%%%%%%%%%%%%%%%%%%%%%%%%%%%%%%%%%%%%%%%%%%%%%%%%%%%%%%%%%%%%%%%%%%%%%%%%%%%%%%%%

%%%%%%%%%%%%%%%%%%%%%%%%%%%%%%%%%%%%%%%%%%%%%%%%%%%%%%%%%%%%%%%%%%%%%%%%%%%%%%%%%%%%%%%%%%%%%%%%%%%%%%%%%%%%
%                                     Table VI                                                             %
%%%%%%%%%%%%%%%%%%%%%%%%%%%%%%%%%%%%%%%%%%%%%%%%%%%%%%%%%%%%%%%%%%%%%%%%%%%%%%%%%%%%%%%%%%%%%%%%%%%%%%%%%%%%
\begin{table}[t]
\caption{$\Omega^{*}$ with spin parity $1/2^{-}$ and $3/2^{-}$ predicted in
the non-relativistic quark model (OGE), relativized quark model
(ROGE), Glozman-Riska model (GBE), covariant quark model based on Bethe-Salpeter
equation with instanton-induced quark force (BSE), large $N_c$ analysis,
algebraic model (BIL), and Skyrme model. The mass is given in the unit of MeV. \label{omega}}
\centering
\begin{tabular}{cccccccc} \hline\hline
State & OGE \cite{CIK81} & ROGE \cite{CI86} & GBE \cite{GR96b} &BSE\cite{metsch2}&
Large-$N_c$ \cite{CC00,SGS02,GSS03,MS04b,MS06b} & BIL \cite{BIL00} &
Skyrme Model\cite{oh} \\
\hline
$\Omega(\frac12^-)$ & $2020$ & $1950$ & $1991$ & 1992 &$2061$ & $1989$ & 1837 \\
                    &        & $2410$ &        & 2456 &       &        &      \\
$\Omega(\frac32^-)$ & $2020$ & $2000$ & $1991$ & 1976 &$2100$ & $1989$ & 1978 \\
                    &        & $2440$ &        & 2446 &       &        & 2604 \\ \hline

\hline\hline
\end{tabular}
\end{table}
%%%%%%%%%%%%%%%%%%%%%%%%%%%%%%%%%%%%%%%%%%%%%%%%%%%%%%%%%%%%%%%%%%%%%%%%%%%%%%%%%%%%%%%%%%%%%%%%%%%%%%%%%%%%%%%%

\newpage
%%%%%%%%%%%%%%%%%%%%%%%%%%%%%%%%%%%%%%%%%%%%%%%%%%%%%%%%%%%%%%%%%%%%%%%%%%%%%%%%%%%%%%%%%%%%%%%%%%%%%%%
%                              Table VII                                                             %
%%%%%%%%%%%%%%%%%%%%%%%%%%%%%%%%%%%%%%%%%%%%%%%%%%%%%%%%%%%%%%%%%%%%%%%%%%%%%%%%%%%%%%%%%%%%%%%%%%%%%%%
\begin{table}[t]
\caption{Energies of the studied five-quark configurations
in the three models with values for hyperfine interactions 
parameters listed in Table~\ref{para}
changed by $\pm20\%$, compared to the results obtained in~\cite{Helminen}
listed in column HR. The results are in unit of MeV.
\label{mass2}}
\renewcommand\tabcolsep{0.88cm}
\renewcommand{\arraystretch}{1}

\vspace{0.3cm}

\begin{tabular}{cccccc} \hline\hline

$J^P$ & OGE &  GBE  & INS  &   HR \\

\hline

${1 \over 2}^-$

& 2120 - 2173   &   1806 - 1862  & 1705 - 1888  & 1917\\

& 2417 - 2485   &   1878 - 1910  & 1982 - 2073  & 1973\\

& 2597 - 2698   &   2005 - 2020  & 2215 - 2232  & 2141\\

&  2611 - 2757  &   2155 - 2168  & 2424 - 2438  & 2381\\

${3 \over 2}^-$

& 1726 - 1910   &  1878 - 1910   &  1934 - 2041 & 1973\\

& 2321 - 2342   &  1987 - 1993   & 1978 - 2072  & 2085\\

& 2458 - 2533   &  2005 - 2020   & 2104 - 2180  & 2141\\

& 2501 - 2548   &   2155 -2168  & 2640 - 2795 & 2381\\

${5 \over 2}^-$

& 2450 - 2535  &   1987 - 1993   & 1934 - 2041  & 2085\\

\hline
\hline
\end{tabular}
\end{table}
%%%%%%%%%%%%%%%%%%%%%%%%%%%%%%%%%%%%%%%%%%%%%%%%%%%%%%%%%%%%%%%%%%%%%%%%%%%%%%%%%%%%%%%%%%%%%%%%%%%%%%%%%%%%%%%%

\end{document}